\documentstyle[pra,aps,twocolumn]{revtex}
\begin{document}
\flushbottom
\draft
\title{
{\bf Phase dynamics in a binary-collisions atom laser scheme}}
\author{O. Zobay and P. Meystre}
\address{Optical Sciences Center, University of Arizona,
Tucson, AZ 85721}
\maketitle
\begin{abstract}
Various aspects of the phase dynamics of an atom laser scheme
based on binary collisions are investigated. Analytical estimates
of the influence of elastic atom-atom collisions on the
laser linewidth are given, and linewidths achievable in a recently
proposed atom laser scheme [Phys.\ Rev.\ A {\bf 56}, 2989 (1997)] are
evaluated explicitly. The extent to which
a relative phase can be established between two interfering atom lasers,
as well as the properties of that phase, are also investigated.
\end{abstract}
\pacs{PACS numbers: 42.50.Vk,03.75.Fi,32.80.Pj,42.50.Ct}
\narrowtext

\section{Introduction}

The recent advances in the generation and manipulation of samples of
ultracold atoms have stimulated research aimed at the development of
sources of coherent atomic beams. Recent MIT experiments
\cite{AndTowMie97,MewAndKur97} can be considered as the first demonstration
of a pulsed atom laser. An important further goal is to achieve a cw system.
Several theoretical proposals to realize such a device have already been
made. They can be separated into two categories:  in the first one,
the population of the atom laser mode results from optical cooling-like
processes such that electronically excited atoms undergo a transition to the
atom laser mode via spontaneous emission
\cite{WisCol95,SprPfaJan95,OlsCasDal95,JanWil96,MoyHopSav97}. Proposals in
the second category
\cite{GuzMooMey96,WisMarWal96,HolBurGarZol96,MooMey971,MooMey972}
rely on binary collisions between ground state atoms fed into an
intermediary level of the atom resonator. These collisions are such that
one of the atoms undergoes a transition into the laser mode and the other
into a heavily damped mode. This latter atom rapidly
escapes the cavity, thereby providing the irreversibility of the
pumping process.

Amongst of the most important characteristics of a laser are the coherence
properties of its output. In the context of cw atom lasers, this aspect has
been addressed in a number of articles:
Refs.~\cite{BalBurSco97,Hop97,NarSchWal97,MoySav97} investigate
various output coupling schemes and their effect on the laser linewidth.
They show that a narrow linewidth can be achieved by a suitable choice of
outcoupling setup. However, the details of the
internal dynamics of the atom laser have not been taken into account so
far in that work. In contrast, Refs.~\cite{WisMarWal96,HolBurGarZol96}
consider the correlation functions for a three-mode model of a
binary-collision atom laser, modeling the output mechanism simply as
linear damping. These studies show that the elastic atomic collisions
which occur in addition to the pumping collisions impose severe limitations
on the coherence of the laser output. Finally, estimates of the linewidth
of an atom laser based on optical cooling are given in Ref.~\cite{WisCol95}.

The purpose of the present article is to extend
the previous work on the phase dynamics of a binary collision atom
laser. We describe this system in terms of a three-mode scheme which ignores
most of the multitude of atomic cavity levels and takes into account only
those modes that are essential for the lasing process, i.~e.\ the pump,
laser mode and loss mode. The feeding and loss processes are described by
linear coupling to external continua.

Several reasons motivate our interest in the binary
collision atom laser and in this particular model.
First, and in contrast to the optical pumping approach, evaporative
cooling has already demonstrated its value in achieving BEC
\cite{AndEnsMat95,DavMewAnd95,BraSacTolHul95}. The binary collision
model we consider can be seen as the simplest possible caricature of
this mechanism. Second, an interesting proposal has recently been made
to overcome some of the problems associated with elastic binary
collisions \cite{MooMey971,MooMey972} in case they result
from the dipole-dipole interaction between atoms: By taking advantage of
specifically designed optical cavities, it is possible to greatly
decrease the rate of elastic collisions between laser mode atoms. It is of
interest to examine the resulting influence on the laser
linewidth. Third, the binary collision atom lasers can be analyzed
within the framework of few-mode models in more detail and more easily
than the models based on optical cooling. An additional
justification for studying such rudimentary models lies in the fact that
they give a good theoretical understanding of the fundamental way an atom
laser might generically work. As such, they provide valuable complementary
insights to more realistic approaches, e.g.\ the quantum kinetic theory of
BEC which may also be applied to atom lasers
\cite{GarZol97,JakGarZol97,GarZolBalDav97}. Finally, before including the
effects of more sophisticated output coupling schemes it is important to
gain a thorough understanding of the internal laser dynamics and its impact
on the laser linewidth in simple models.

After briefly reviewing the main aspects of binary collisions atom laser
schemes in section II, the present paper addresses two main topics:
Section III discusses the laser linewidth and its
dependence on elastic atom-atom collisions. In previous studies
this question was either treated rather briefly \cite{HolBurGarZol96}
or estimates were given \cite{WisMarWal96} the validity of which
is not quite clear \cite{Wis97}. Our analysis leads to an analytical
approximation for the laser linewidth in the presence of interatomic
collisions. To this end a linearized fluctuation analysis of the system
master equation is performed after adiabatically eliminating the loss
mode. The results of this analysis are compared to quantum Monte Carlo
simulations \cite{DumParZolGar92,MolCasDal93}. We illustrate these results
in the determination of the possible operating regimes
of the atom laser of Refs.~\cite{MooMey971,MooMey972}, and demonstrate that
in for weak pumping it should in principle be possible to obtain
laser linewidths below the natural linewidth of the laser mode.

In section IV we turn to the question of whether a definite relative phase
can be established between independent atom lasers. The same question, but
for Bose-Einstein condensates, has aroused much interest recently
\cite{JavYoo96,NarWalSch96,CirGarNar96,WonColWal96,CasDal97,JavWil97}.
It was shown that for a system of two condensates with a fixed total number
of atoms, a relative phase is established by the process of measurement.
Our analysis extends this work to the realm of open systems. We show
that a relative phase can be established in single runs of the
experiment in very much the same way as for Bose condensates.
The study of the diffusion properties of this relative phase shows its
close connection to the phase of the single atom laser as described in
terms of stochastic processes. Finally, a brief summary and conclusions
are given in section V.

\section{The binary collision atom laser model}

The basic principles of operation of the binary collision atom laser
are discussed in detail in several publications
\cite{GuzMooMey96,WisMarWal96,HolBurGarZol96,MooMey971,MooMey972}.
One considers a resonator for atoms, realized e.g. by optical fields.
In order to concentrate on the essential dynamics only three out of the
multitude of atomic center-of-mass modes are taken into account explicitly
(cf.\ Fig.~\ref{fig21}). Bosonic atoms in their ground electronic state are
pumped into an atomic resonator level of ``intermediary'' energy (mode 1).
They then undergo binary
collisions which take one of the atoms involved to the tightly bound
laser mode 0, whereas the other one is transferred to the heavily damped
loss mode 2. This latter atom leaves the resonator quickly, thereby
providing the irreversibility of the pumping process. A macroscopic
population of the laser mode can built up as soon as the influx of
atoms due to pumping compensates for the losses induced by the damping.

In the description of this laser scheme one has to take into
account that in addition to the pumping collisions other types of
interatomic collisions can also occur. These considerations lead to an
ansatz for the atom laser master equation of the form
\begin{eqnarray} \label{al3m}
\dot{W}& = & -i[H_0+H_{col},W]+\kappa_0{\cal D}[a_0]W+\kappa_1(N+1){\cal D}
[a_1]W \nonumber \\
 & & +\kappa_1 N {\cal D}[a_1^{\dagger}]W + \kappa_2{\cal D}[a_2]W
\end{eqnarray}
with $\hbar=1$.
In this equation, we use the second quantized formalism in which each
center-of-mass atomic mode is associated with an annihilation operator
$a_i$, and $W$ denotes the atomic density operator\footnote{Since we
consider ground state atoms only, they
are fully described by their center-of-mass quantum numbers.}.
The free Hamiltonian is given by
$$H_0=\sum_{i=0,1,2} \omega_i a_i^{\dagger} a_i,$$
$\omega_i$ being the mode frequencies. The general form of the collision
Hamiltonian can be written as
\begin{equation}\label{hcol}
H_{col}=\sum_{i\leq j,k\leq l} V_{ijkl} a^{\dagger}_i a^{\dagger}_j
a_k a_l
\end{equation}
with $V_{ijkl}$ the matrix elements of the two-body interaction
Hamiltonian responsible for the collisions. However, for the present
purposes it will be sufficient to restrict our attention to the reduced form
\begin{eqnarray}\label{hcolres}
H_{col} & = & V_{0211}a^{\dagger}_0 a^{\dagger}_2 a_1 a_1 + V_{1102}
a^{\dagger}_1 a^{\dagger}_1 a_0 a_2 + V_{0000} a^{\dagger}_0
a^{\dagger}_0 a_0 a_0 \nonumber \\
 &  & + V_{0101} a^{\dagger}_0 a^{\dagger}_1 a_0 a_1 +
V_{1111} a^{\dagger}_1 a^{\dagger}_1 a_1 a_1
\end{eqnarray}
in which (besides the pumping collisions) only those collisions are
retained which are expected to have the most significant influence on
the phase dynamics. The damping rates of the
cavity modes are given by the coefficients $\kappa_i$, and the strength of
the external pumping of mode 1 is characterized by the parameter $N$,
which is the mean number of atoms to which mode 1 would equilibrate in
the absence of collisions. The superoperator ${\cal D}$ is of the Lindblad
form and is defined by
\begin{equation}
{\cal D}[c] P = cPc^{\dagger}-\textstyle{\frac 1 2}(c^{\dagger} c P -
P c^{\dagger} c)
\end{equation}
with arbitrary operators $c$ and $P$.

It has been pointed out \cite{WisMarWal96} that it is useful to
study this system in detail although it might be thought of more as a
caricature than as an approximation, due to the neglect of most of the
modes in the atomic resonator. From this point of view, its main purpose is
to give a basic understanding of the laser dynamics, rather than a detailed
quantitative description of a realistic setup. However, recent work has
added significantly more substance to this model \cite{MooMey971,MooMey972}:
It was shown that in case the binary collisions result from the
dipole-dipole interaction and for a quasi one-dimensional modulated
cavity, a high mode selectivity for the atomic collisions can be obtained.
One particular pumping mode is then coupled only to a small number
of other modes. Among the small set of initially populated low-lying modes,
a particular one is then expected to be singled out as a result of mode
competition. Another benefit of this model is the fact that due to the
dipole-dipole selection rules in the resonator, the strength $V_{0000}$ of
detrimental collisions can be made
small in comparison to $V_{0211}$. In that proposal, an efficient pumping
process is achieved by a time-dependent modulation of the atom cavity. In
the spirit of this model we set $\omega_0 + \omega_2 = 2\omega_1$ in the
following.

In order to achieve a sufficiently high degree of irreversibility it is
necessary that $\kappa_2$ is much larger than the damping rates of the
other modes. This suggests to adiabatically eliminate
this mode, an approximation that leads to the simplified master equation
\cite{GuzMooMey96,WisMarWal96,HolBurGarZol96}
\begin{eqnarray} \label{al2m}
\dot{\rho} & = & -i[H_{c},\rho]+\kappa_0{\cal D}[a_0]\rho
+\kappa_1(N+1){\cal D} [a_1]\rho \nonumber \\
 & & +\kappa_1 N {\cal D}[a_1^{\dagger}]\rho + \Gamma {\cal D}
[a_0^{\dagger}a_1^2]\rho.
\end{eqnarray}
Equation (\ref{al2m}) is written in the interaction picture with respect
to $H_0=\omega_0 a^{\dagger}_0 a_0 + \omega_1 a^{\dagger}_1 a_1$, and the
reduced density matrix $\rho$ is
$\rho=\mbox{Tr}_{\mbox{mode 2}}[W]$. The reduced collision Hamiltonian
$H_c$ is
\begin{equation}
H_c = V_{0000}a_0^\dagger a_0^\dagger a_0 a_0 + V_{0101}a^{\dagger}_0
a^{\dagger}_1 a_0 a_1 + V_{1111}a_1^\dagger a_1^\dagger a_1 a_1 ,
\end{equation}
and $\Gamma=4|V_{0211}|^2/\kappa$. Consistently
with Ref.~\cite{WisMarWal96} we call the limiting cases
$\Gamma \ll \kappa_0$ and $\Gamma \gg \kappa_0$ the weak and strong
collision regimes, respectively. The reduced master equation (\ref{al2m})
forms the basis of most previous studies of binary collision atom lasers.

The master equations (\ref{al3m}) and (\ref{al2m}) can easily be solved
numerically using standard quantum Monte Carlo simulation techniques
\cite{DumParZolGar92,MolCasDal93}. This is discussed in detail
in Ref.~\cite{MooMey972}, and we only recall that these simulations
are significantly facilitated by the fact that between quantum jumps the
total number of atoms is conserved. In this way one has to propagate only
a small number of actually populated states, a situation evidently
advantageous with regard to both memory and CPU time
requirements\footnote{In particular, in the calculation of
correlation functions of the type $\langle a^{\dagger}_j(t+\tau)a_j(t)
\rangle,$ one normally has to evolve four wave functions $[1\pm (i) a_j]
|\Phi(t)\rangle$ for a given initial sample wave function $|\Phi(t)\rangle$
if one uses the method of Ref.~\cite{MolCasDal93}.
However, due to the conservation of atom number in this case
the contributions from $|\Phi(t)\rangle$ and $a_j|\Phi(t)\rangle$
in the time development of $[1 + a_j]|\Phi(t)\rangle$ can be distinguished.
This means that one obtains a contribution to the correlation function even
from a single propagation. As it is more advantageous
to average over a large number of different $|\Phi(t)\rangle$ than to
use many simulations for a small number of initial wave functions one
obtains a significant additional decrease in computation time.}.

\section{Elastic collisions and laser linewidth}

\subsection{Linearized fluctuation analysis for the two-mode system}

In order to obtain an analytical approximation for the laser linewidth
in the two-mode sytem a linearized fluctuation analysis can be performed
\cite{MeySar91,WalMil94}.
To this end the master equation (\ref{al2m}) is converted to a
Fokker-Planck equation using the $P$-function representation as described
in \cite{WisMarWal96}. This equation can be transformed to polar
coordinates $\alpha_j=\sqrt{n_j}e^{i\phi_j}$, where $\alpha_j$ denotes the
complex amplitudes originally appearing in the Fokker-Planck equation
\cite{Gar83,Gar91}. This leads to stochastic differential equations
\begin{eqnarray}
dn_0&=&[\Gamma n_1^2 (n_0+1) - \kappa_0 n_0] dt + dS_{n_0}, \label{sden0} \\
dn_1&=&[\kappa_1(N-n_1)-2\Gamma n_1^2(n_0+1)]dt + dS_{n_1}, \label{sden1} \\
d\phi_0&=&[-V_{0000}(2n_0-1) -V_{0101}n_1]dt + dS_{\phi_0}, \label{sdephi0} \\
d\phi_1&=&[-V_{1111}(2n_1-1) -V_{0101}n_0]dt + dS_{\phi_1}. \label{sdephi1}
\end{eqnarray}
The correlation matrix for the stochastic forces $d{\bf S}^T=(dS_{n_0},
dS_{n_1},dS_{\phi_0},dS_{\phi_1})$ is given by
\begin{equation} \label{diffmat}
{\bf D}=\hspace*{-2mm}
\left(\begin{array}{cccc} 2\Gamma n_1^2 n_0 & -2\Gamma n_1^2 n_0
& -2V_{0000}n_0 & -V_{0101} n_0 \\
-2\Gamma n_1^2 n_0 & 2\kappa_1 Nn_1-2\Gamma n_1^2 n_0 & -V_{0101}n_1 &
-2V_{1111}n_1 \\
-2V_{0000}n_0 & -V_{0101}n_1 & \Gamma n_1^2/(2n_0) & \Gamma n_1/2 \\
-V_{0101} n_0 & -2V_{1111}n_1 & \Gamma n_1/2 & \frac{\kappa_1 N}
{2n_1} + \frac{\Gamma n_0}{2} \end{array} \right).
\end{equation}
In the limit $n_0\gg 1$ one obtains from Eqs.~(\ref{sden0})
and (\ref{sden1}) the above-threshold semiclassical steady-state
populations \cite{WisMarWal96}
\begin{eqnarray}
\bar{n}_0&=&\frac 1 2 \frac{\kappa_1}{\kappa_0}(N-\bar{n}_1),\label{ssn0} \\
\bar{n}_1&=&\sqrt{\frac{\kappa_0}{\Gamma}}, \label{ssn1}
\end{eqnarray}
the threshold condition being $N > \sqrt{\kappa_0/\Gamma}$.
The relation (\ref{ssn0}) between $\bar{n}_0$ and $\bar{n}_1$ also holds
in the full quantum-mechanical two- and three-mode models \cite{MooMey972}.
To proceed further we introduce the fluctuation variables
$\delta n_j=n_j-\bar{n}_j$ and $\delta \phi_j=\phi_j - \bar{\phi}_j$, where
the phase drift variables $\bar{\phi}_j$ obey the deterministic equations
obtained from Eqs.~(\ref{sdephi0}) and (\ref{sdephi1}) by discarding
the stochastic forces and substituting $\bar{n}_j$ for $n_j$.
In the linear approximation the time evolution of the fluctuations
$\delta {\bf A}^T=(\delta n_0,\delta n_1,\delta \phi_0,\delta \phi_1)$ is
given by
\begin{equation}
d\,\delta {\bf A}= - {\bf k}\, \delta {\bf A} dt + d{\bf S}
\end{equation}
where the matrix ${\bf k}$ is obtained by linearizing the drift terms in
Eqs.~(\ref{sden0}) -- (\ref{sdephi1}) around the steady-state values
$\bar{n}_j$. The correlation matrix for the stochastic forces $d{\bf S}$
is given by the matrix ${\bf D}$ of Eq.~(\ref{diffmat}) after replacing
$n_j$ by $\bar{n}_j$. Supposing that $\delta {\bf A}={\bf 0}$ at time
$\tau=0$, the probability distribution of the fluctuations at a later
time $\tau$ is given by \cite{Gar83}
\begin{equation}
p(\delta {\bf A},\tau)=[(2\pi)^4 \det {\bbox \sigma}(\tau)]^{-1/2} \exp
[-\textstyle{\frac 1 2}\delta {\bf A}^T {\bbox \sigma}^{-1}(\tau)\delta
{\bf A}]
\end{equation}
with
\begin{equation}\label{defsigma}
{\bbox \sigma}(\tau)=\int_0^{\tau} d\tau' \exp[-{\bf k}(\tau-\tau')]\,
{\bf D} \exp[-{\bf k}^T(\tau-\tau')].
\end{equation}
Assuming as usual that the first-order correlation function $C_0(\tau)=
\langle a_0^{\dagger}(\tau) a_0(0) \rangle$ is determined only by the
phase fluctuations one obtains
\begin{eqnarray}\label{defc0}
C_0(\tau)&=& \int d^4 \delta {\bf A} \,\bar{n}_0 e^{-i \phi_0} p(\delta
{\bf A},\tau) \nonumber \\
&=& \bar{n}_0 e^{-i \bar{\phi}_0} \exp[-\textstyle{\frac 1 2}
\sigma_{33}(\tau)] \label{c0},
\end{eqnarray}
where the deterministic phase drift is also included.
This phase drift leads to a shift of the center of the power
spectrum (the Fourier transform of the correlation function) by an
amount $2V_{0000}\bar{n}_0+V_{0101}\bar{n}_1$ with respect to the
collisionless case. The behavior of the correlation function $C_0(\tau)$
is thus essentially determined by the covariance matrix element
$\sigma_{33}(\tau)$.

The right hand side of Eqs.~(\ref{sden0}) -- (\ref{sdephi1}) depends only
on the atom numbers $n_0$ and $n_1$\footnote{This is in
contrast to the three-mode system, where an explicit
phase dependence is introduced through the coherent
pumping term $V_{0211}a_0^{\dagger}a_2^{\dagger}a_1 a_1 + h.c.$}.
This means that in that two-mode scheme the phase
diffusion due to interatomic collisions is induced solely by atom number
fluctuations. This observation significantly simplifies the further
analytical treatment as it leads to the matrix ${\bf k}$ having two vanishing
column vectors. Hence the covariance matrix ${\bbox
\sigma}(\tau)$ can be evaluated explicitly in a straightforward way
according to Eq.~(\ref{defsigma}). However, the ensuing expression is
still rather complicated, so that in the following we restrict the
discussion to two limiting cases which illustrate the essential aspects
of the influence of the elastic collisions on the laser linewidth.
It should also be noted at this point that in the two-mode system
$C_0(\tau)$ does not depend on elastic collisions between pumping mode
atoms, which are characterized by the parameter $V_{1111}$.

({\it a}) $V_{0101}=0$. Expanding the expression for $\sigma_{33}(\tau)$
to leading order in the parameter $\bar{n}_0$ one obtains
\begin{eqnarray}\label{sigma33}
\sigma_{33}(\tau)&=& \tau [w + \kappa_0/(2\bar{n}_0)] \nonumber \\
& & +\frac{[1-\exp(-q\tau)]}{q} (8V_{0000}^2\bar{n}_0/q - 2w)
 \nonumber \\
& & +\frac{[1-\exp(-2q\tau)]}{2q} (w-8V_{0000}^2\bar{n}_0/q)
\end{eqnarray}
with
\begin{equation}
w=V_{0000}^2\frac{\kappa_1 N}{\kappa_0^2} \left(2+2\sqrt{\kappa_0/
\Gamma}+\frac{N}{N-\sqrt{\kappa_0/\Gamma}}\right)
\end{equation}
and
\begin{equation}\label{defq}
q=\frac{4\kappa_0^2}{4\kappa_0+\kappa_1 \bar{n}_1/\bar{n}_0}.
\end{equation}
>From Eq.~(\ref{ssn0}) it follows that $\kappa_1 N$ and thus $w$ are of the
order of $\bar{n}_0$. In the first line of Eq.~(\ref{sigma33}) the term
$\kappa_0/(2\bar{n}_0)\tau$ was included although it is not of the same
order in $\bar{n}_0$ as the other terms. This is because it describes
the behavior of the correlation function if $V_{0000}$ is very small
(cf.\ Fig.~\ref{fig32}). The parameter $q$ of Eq.~(\ref{defq}) is one of
the eigenvalues of the upper left $2\times 2$-minor of the matrix ${\bf
k}$. It can thus be interpreted as the inverse of one of the timescales
relevant in the dynamics of the atom number fluctuations. The other
eigenvalue is of the order of $\bar{n}_0$ and introduces a much shorter
time scale. It does not play a significant role for the characterization of
the correlation function.

Equation (\ref{defc0}) implies that the most important aspects of the
correlation function can be inferred from the study of the behavior of
$\sigma_{33}(\tau)$ in the time interval where it is smaller than or of the
order of unity. From Eq.~(\ref{sigma33}) two different kinds of
behavior can thus be distinguished, depending on
the magnitude of $V_{0000}$. As long as $\sigma_{33}(\tau =1/q)\ll 1$
the time evolution of $\sigma_{33}(\tau)$ relevant for $C_0(\tau)$
is well approximated by
\begin{equation}\label{g00small}
\sigma_{33}(\tau)\simeq \tau [w + \kappa_0/(2\bar{n}_0)].
\end{equation}
This is because as time increases the second and the third term on the
right-hand side of Eq. (\ref{sigma33}) remain constant for $\tau > 1/q$,
while the first term
increases. For such values of $V_{0000}$ $C_0(\tau)$ decays therefore
exponentially, and the power spectrum is Lorentzian. If $w \gg \kappa_0/
(2\bar{n}_0)$ the linewidth is proportional to $V_{0000}^2\bar{n}_0$. IN
contrast to the situation with conventional lasers, it {\em increases}
linearly with the number of atoms in the laser mode.

In case $\sigma_{33}(\tau=1/q)\gg 1$ the decay of the correlation function
can accurately be approximated by expanding the exponentials in
Eq.~(\ref{sigma33}) up to second order. This yields the expression
\begin{equation}\label{g00large}
\sigma_{33}(\tau)\simeq4V_{0000}^2\bar{n}_0\tau^2.
\end{equation}
Under these circumstances the correlation function decays like a Gaussian.
The spectrum is thus itself a Gaussian and its linewidth is proportional
to $V_{0000}\sqrt{\bar{n}_0}$. The atom laser linewidth still increases
with ${\bar n}_0$, albeit less dramatically than in the preceding case.

Figure~\ref{fig32} suggests that a rough
estimate of the value of $V_{0000}$ at which the decay of the correlation
function changes from exponential to Gaussian can be obtained in the
following way. One determines $V_{0000}$ such that both
Eqs.~(\ref{g00small}) and (\ref{g00large}) yield the same solution for
the condition  $\sigma_{33}(\tau)=2/\ln 2$. In this way one obtains as
``critical'' value
\begin{equation}\label{V00crit}
V_{0000,\, crit}=\sqrt{\frac{8}{\ln 2}}\frac{\sqrt{\bar{n}_0}}{w'}
\end{equation}
with $w'=w/V_{0000}^2$. If $\bar{n}_0$ is increased the change thus
occurs for smaller values of $V_{0000}$.

It is instructive to compare these results to those obtained for a damped
harmonic oscillator with Hamiltonian $H=\omega_0 a_0^{\dagger}a_0 +
V_{0000} a_0^{\dagger}a_0^{\dagger}a_0 a_0$. Denoting in analogy with
Eq.~(\ref{al3m}) the damping coefficient by $\kappa_0$ and the external
pumping strength by $N$, a linearized fluctuations analysis yields for
the relevant covariance matrix element
\begin{eqnarray}\label{harmosc}
\bar{\sigma}(\tau)&=&\tau\left[8V_{0000}^2\frac{N^2+N}{\kappa_0}
+\kappa_0/2\right] \nonumber \\
& & -16V_{0000}^2\frac{N^2+N/2}{\kappa_0^2}\left[1-\exp(-\kappa_0 \tau)
\right] \nonumber \\
& & +4V_{0000}^2\frac{N^2}{\kappa_0^2}\left[1-\exp(-2\kappa_0 \tau) \right],
\end{eqnarray}
where no further approximation has been made, in contrast to the case of
Eq.~(\ref{sigma33}). In the limit of large $V_{0000}$ Eq.~(\ref{harmosc})
yields $\bar{\sigma}(\tau)=4V_{0000}^2 N\tau^2$, a result formally identical
to Eq.~(\ref{g00large}) since the equilibrium population $\bar{n}_0$
becomes $N$ in the one-mode system.  Such a connection between the linewidths
for this one-mode system and the atom laser has been previously noted in
Ref.~\cite{Wis97}. Despite their formal similarity, there are however
significant differences between Eqs.~(\ref{sigma33}) and (\ref{harmosc}).
First, there is no obvious correspondence between the coefficients
multiplying their respective time-dependent contributions, 
and they are not of the same
order in $\bar{n}_0$. In addition, the one-mode system is characterized by
a very large ``fundamental'' linewidth $\kappa_0/2$ which masks a possible
quadratic dependence on $V_{0000}$. Finally, the above-mentioned limiting
behavior is reached for values of $V_{0000}$ much larger than in the 
two-mode atom laser system.

({\it b}) $V_{0000}=0$. In this case the expansion of $\sigma_{33}(\tau)$
to leading order in $\bar{n}_0$ yields
\begin{equation}\label{g01app}
\sigma_{33}(\tau)=\left(\frac{V_{0101}^2}{2\Gamma \bar{n}_0}+
\frac{\kappa_0}{2\bar{n}_0}\right) \tau
\end{equation}
The correlation function thus decays exponentially for all values of
$V_{0101}$. A qualitative change in behavior as in the previous
situation does not occur. The linewidth of the spectrum is now proportional
to $V_{0101}^2/\bar{n}_0$, and becomes {\em narrower} when the 
population of the laser mode is increased, very much like
the familiar Shawlow-Townes linewidth of conventional lasers.

The analytical results described so far were compared to numerical
calculations of the correlation function. Both the weak-collision
regime and the strong-collision regime were investigated for the explicit
values $\Gamma/\kappa_0=1/15$ and $15$ and
$\kappa_1/\kappa_0=10$ and 100. For $V_{0101}=0$ a good agreement could
generally be observed between the numerical result and the
analytical approximation based on Eq.~(\ref{defc0}). A representative
example is shown in Fig.~\ref{fig31} where the parameter values are
$\Gamma/\kappa_0=15$, $\kappa_1/\kappa_0=100$, $N=1.85$,
$V_{0000}/\kappa_0=2.5$, $V_{0101}=0$. The numerical results are well
approximated by a Gaussian. Considerable deviations occurred only in
the case $\Gamma/\kappa_0=1/15$, $\kappa_1/\kappa_0=100$. For these values
the quantum effects in the
population dynamics are particularly large, e.g.\ choosing $N$ such
that $\bar{n}_0=50$ according to Eq.~(\ref{ssn0}) led to a numerical
value of 100 whereas in the other case deviations were only a few percent.
It should be noted that in the two-mode systen the population dynamics do
not depend on the interatomic collisions.

The transition between the exponential and
Gaussian regimes in the behavior of the correlation function
$C_0(\tau)$ is illustrated in Fig.~\ref{fig32}. It shows
the inverse of the ``half-life time'' $\tau_{1/2}$ of $C_0(\tau)$
(defined by $\sigma_{33}(\tau_{1/2})=\ln 2/2$) as a
function of $V_{0000}$ for the parameters $\Gamma/\kappa_0=15$,
$\kappa_1/\kappa_0=10$, and $N=20.3$, corresponding to $\bar{n}_0=100$,
on a doubly-logarithmic scale. The full curve depicts $\tau_{1/2}^{-1}$
as determined from Eq.~(\ref{defc0}), the dashed and the dotted
curves are obtained from the approximate relations (\ref{g00small}) and
(\ref{g00large}), respectively. The results of numerical Monte Carlo
simulations are shown as circles ($\bullet$). The figure indicates that
the analytical predictions, in particular the existence of two different
regimes and the transition between them, are accurately confirmed by the
numerical calculations. The approximations based on Eqs.~(\ref{g00small})
and (\ref{g00large}) are almost indistinguishable from the full analytical
expression in their respective regions of validity.

The results of a similar study for the collision coefficient $V_{0101}$
are shown in Fig.~\ref{fig33}. There, the dashed curve is derived from
the approximation (\ref{g01app}). The global behavior of the numerical
results is well approximated by the analytical description, in particular
the quadratic dependence of the half-life time on $V_{0101}$ is
recovered. However, the analytical prediction typically overestimates
the numerical half-life time by a factor of two. This behavior is also
observed in examples with other parameter values. This indicates that the
atom number fluctuations in the pump mode are not as well described by a
linearized ansatz as the fluctuations in the laser mode.

In case both collision coefficients $V_{0000}$ and $V_{0101}$ are
non-vanishing the analytical form of a reliable approximation to
Eq.~(\ref{defsigma}) would be rather complicated. However, it can be
seen that all terms involving $V_{0101}$ are at most of the order of
$\bar{n}_0^0$. Taking into account Eq.~(\ref{V00crit}) this means that
in the limit of large $\bar{n}_0$ the function $\sigma_{33}(\tau)$ can
eventually be approximated by Eq.~(\ref{g00large}). The comparison with
numerical results shows that Eq.~(\ref{defc0}) is still accurate when both
collision coefficients are present. The calculations indicate that the
correlation time decreases, i.e., the linewidth broadens, if the second
collision coefficient is switched on in addition to the first. If the
values of the collision coefficients are such that one of them has a much
larger impact on the correlation time than the other -- which is checked
by comparing Eqs.~(\ref{sigma33}) and (\ref{g01app}) -- then the
approximate results of the limiting cases ({\it a}) and ({\it b}) still hold.
If both have roughly the same impact then the ensuing correlation time
will still be of the order of the result inferred from the approximations.

\subsection{Operating regimes of the atom laser}

In the following we apply these results to the atom
laser scheme of Refs.~\cite{GuzMooMey96,MooMey971,MooMey972}. In this
model the collision mechanism is the laser-induced dipole-dipole
interaction. Due to energy and momentum conservation, the associated
selection rules allow certain collision coefficients, most notably 
$V_{0000}$, to become small in comparison with $V_{0211}$. In our
numerical example we examine a situation where only the elastic collision 
coefficients $V_{0000}=V_{0211}/15$ and $V_{0101}=V_{0211}$ are taken into 
account. A further interesting feature of this system is the fact that 
the dominant collision mechanism and the atom cavity are realized with
laser fields whose intensities can be varied independently from each other.
This means that while the ratio $V_{ijkl}/V_{0211}$ is virtually fixed once
the experimental setup has been optimized there is still considerable
latitude in the choice of the relative magnitude of $V_{0211}$ compared
to $\kappa_0$.

For the numerical simulations involving the three-mode model to remain 
tractable, we chose parameters such that the laser linewidth is
of the order of the natural linewidth $\kappa_0$ and the equilibrium
laser mode population of the order of $\bar{n}_0=50$. With the help of
Eq.~(\ref{g00large}) one estimates from the condition
$\sigma_{33}(2/\kappa_0)=1$ that $V_{0000}=0.035\kappa_0$. This also fixes
$V_{0101}=V_{0211}=15V_{0000}$. From Eq.~(\ref{g01app}) one obtains
$\Gamma=5.63\times 10^{-3}\kappa_0$ as a suitable value for the pumping
parameter by again using the condition $\sigma_{33}(2/\kappa_0)=1$.
Choosing $\kappa_1=20 \kappa_0$ yields $N=18.3$. In this way all
parameters which enter the two-mode model are determined. The
relative orders of magnitude appear to be reasonable. The numerical
calculation yields a half-life time of about $0.5\kappa_0$ and
$\bar{n}_0=68$ which agrees reasonable well with our stated objective.

In order to put this result into the right perspective it is
necessary to perform corresponding calculations for the three-mode
model also. This is because the two-mode scheme assumes that
$V_{0211},\kappa_2 \to \infty$ with the ratio $\Gamma=4V_{0211}^2/\kappa_2$
held fixed, without taking into account the need to assign finite values
to the ratios $V_{ijkl}/V_{0211}$.

One of the most important results of a systematic comparison between the
predictions of the two-mode and the three-mode models is
that in the presence of elastic collisions the equilibrium populations
may react as least as sensitively to the effects of ``non-adiabaticity''
as the linewidth. In other words, as the
values of $V_{0211}$ and $\kappa_2$ are lowered with $\Gamma$ fixed,
$\bar{n}_0$ can drop significantly below its adiabatic value even
before the linewidth is considerably affected. This occurs
as soon as the value of $V_{0211}$ becomes of the order of the other
collision coefficients. This may appear surprising in view of the fact that
in the two-mode system the population dynamics are completely independent of
the elastic collisions. A possible explanation
is that the levels $|n_0,n_1,n_2\rangle$ and $|n_0+1,n_1-2,n_2+1
\rangle$ which are coupled by the coherent pumping experience an
effective detuning due to the collisions. Hence, the predictions
of the two-mode system for a given set of parameters should always be
compared the full three-mode simulations to assess their
reliability. In the present example, however, the population decrease in
the three-mode system (with $\kappa_2=4V_{0211}^2/\Gamma =201\kappa_0$)
is not very pronounced ($\bar{n}_0=48$) while the linewidth hardly changes
compared to its two-mode value. These results indicate that in the atom
laser scheme of Refs.~\cite{GuzMooMey96,MooMey971,MooMey972} it should
be possible to obtain a linewidth of the order of or smaller than the
natural linewidth $\kappa_0$ of the laser mode.

\section{Relative phase between two atom lasers}

The question of how a definite relative phase can be established
between two independently created, interfering Bose condensates has been
the subject of intensive theoretical study in the recent past
\cite{JavYoo96,NarWalSch96,CirGarNar96,WonColWal96,CasDal97,JavWil97}.
It was shown that one needs not resort to the concept of spontaneously
broken symmetry to model the creation of interference patterns.
Rather, these structures are brought about by the very act of observation:
A sequence of measurements leads to the creation of an entangled state of
the condensates pair which can be assigned a relative phase. However, this 
phase varies randomly from one realization of the experiment to the other.

In this section we examine whether such a relative
phase is also created when the output beams of two independent atom lasers
are brought to interference. The purpose of this study is to determine
to which extent the concept of measurements-induced phase built-up
retains its validity also in open systems. The entanglement between the
atoms in the two laser modes is now influenced by the
continuing pumping processes, and it is natural to ask if a well-defined
relative phase can persist and what its properties are, e.g.\ with respect
to diffusion. The question is also of interest with regard to related
studies of optical systems \cite{Mol97}.

We consider a gedanken experiment where the outputs of two
two-mode atom lasers are brought to interference
with the help of a 50:50 atomic beam splitter\footnote{The use of a 
two-mode rather than a three-mode laser model
simplifies the numerics while retaining the essential physical
characteristics of the binary collision model. Furthermore, in the Monte
Carlo wave function of a single two-mode atom laser a single number state
is populated, so that it closely resembles the state of an isolated
Bose condensate.}. In the spirit of Ref.~\cite{CasDal97}, the effective
non-hermitian Hamiltonian describing the Schr\"odinger-like evolution of
the two-laser system is
\begin{equation}\label{Hefftot}
H_{eff}=H_a+H_b
\end{equation}
where
\begin{eqnarray}\label{Heffa}
H_a & = & H_c-\textstyle{\frac i 2}[\kappa_0 a_0^{\dagger}a_0+\kappa_1
(N+1) a_1^{\dagger}a_1 \nonumber \\
& & + \kappa_1 a_1 a_1^{\dagger}+\Gamma a_1^{\dagger\, 2}
a_0 a_0^{\dagger} a_1^2]
\end{eqnarray}
is the effective Hamiltonian for the first atom laser. The Hamiltonian for
the second laser is defined similarly, but using annihilation
operators $b_i$. The definition of the various quantum jump operators is
obvious from Eqs.~(\ref{Hefftot}) and (\ref{Heffa}) except for those
which describe atoms leaving the laser modes after passing through the
beam splitter. They are given by
\begin{equation}
c_{\pm}=(a_0 \pm b_0)/\sqrt{2}.
\end{equation}

It is a simple matter to perform quantum Monte Carlo simulations
within this framework. The general form of the simulation wave function
can be written as
\begin{equation}\label{MCwf}
|\Psi(t)\rangle=\sum_{n_{01}=n_{01,min}}^{n_{01,max}} c_{n_{01}}(t)
|n_{01},n_{11},n_{02}=N_L - n_{01},n_{12} \rangle,
\end{equation}
where $n_{ij}$ denotes the number of atoms in mode $i$ of laser $j$
and $N_L$ is the total number of atoms in the two laser modes.
The distribution function of the relative phase between the
laser modes is conveniently expressed in terms of the overcomplete set
of phase states
\cite{CasDal97}
\begin{equation}\label{phasestate}
|\Phi\rangle_{N_L}=\frac{1}{\sqrt{2^{N_L}N_L!}}(a_0^{\dagger}e^{i\Phi/2}+
b_0^{\dagger}e^{-i\Phi/2})^{N_L}|0,0,0,0\rangle.
\end{equation}
Because of the relation
\begin{equation}\label{cosrel}
(a_0+b_0)|\Phi\rangle_{N_L}=\sqrt{2N_L} \cos (\Phi/2) |\Phi\rangle_{N_L}
\end{equation}
one can think of the state $|\Phi\rangle_{N_L}$ as possessing a definite
relative phase $\Phi$ between the laser modes within the context of the
interference experiment. The wave function $|\Psi(t)\rangle$ can be
expanded on these phase states as
\begin{equation}
|\Psi(t)\rangle=\int_{-\pi}^{\pi} \frac{d\Phi}{2\pi}\, c(\Phi,t) \,
|\Phi\rangle_{N_L}
\end{equation}
where the phase distribution amplitude $c(\Phi, t)$ is given by
\begin{eqnarray}
c(\Phi,t) & = & 2^{N_L/2} \sum_{n_{01}=n_{01,min}}^{n_{01,max}}
\left(\frac{n_{01}! (N_L-n_{01})!}{N_L!}\right)^{1/2} \nonumber \\
& & e^{i(N_L /2 -n_{01})\Phi} c_{n_{01}}(t).
\end{eqnarray}

Several physical processes influence the phase distribution: they are (i)
the loss processes induced by the action of the operators $c_{\pm}$
leading to the entanglement of the atom lasers;
(ii) the pumping into the laser modes
described by the action of $a_{0}^{\dagger}$ and $b^{\dagger}_0$; (iii)
the collision interactions taking place between quantum jumps and described
by $H_c$; and
(iv) the non-Hermitian time evolution described by the $\Gamma$-term in
Eq.~(\ref{Heffa}). Whereas (i) and (iii) are also present in the case of
Bose condensates, the influence of (ii) and (iv) occurs only for the
laser. It is not a priori evident that the distribution function
$c(\Phi,t)$ still yields a well-defined phase in the presence of these
terms. 

Figure \ref{fig41} shows typical phase distributions $|c(\Phi)|^2$ obtained 
from Monte Carlo wave function simulations of the two-laser problem. 
All examples in this section are calculated for the
parameters $\Gamma=21.9\kappa_0$, $\kappa_1=54.6\kappa_0$ and $N=2.2$. 
Figure \ref{fig41}(a) illustrates the collisionless case $V_{0000}=0$.
Obviously, in this case the state of the system can be ascribed a 
well-defined phase. The distribution is
symmetric with respect to $\Phi=0$, reflecting the fact that the method
of detection does not discriminate between the states $|\Phi\rangle$ and
$|-\Phi\rangle$. However, it is much broader than
for the corresponding phase state $|\Phi\rangle_{N_L}$  shown in
Fig.~\ref{fig41}(b) for comparison. That means that in general an
approximation of the instantaneous Monte Carlo wave function by the
superposition of two phase states is not appropriate.

The influence of elastic collisions on the phase distribution is
demonstrated in Fig.~\ref{fig41}(c). There, the value $V_{0000}=0.25\kappa_0$
was chosen, yielding a half-life time of $\tau_{1/2}=0.33/\kappa_0$
for the correlation function of the single atom laser, a value much smaller
than the value $\tau_{1/2}=142/\kappa_0$ for the collisionless
case. Figure \ref{fig41}(c) shows that the collisions tend to broaden
the phase distribution. Furthermore the symmetry of the distribution is
lost with respect to both the height and the center of the two maxima.
Nonetheless it is still reasonable to associate a well-defined phase with
this state.

From a formal point of view it is not too surprising that the Monte
Carlo wave functions retain their definite-phase character in the
presence of the pump processes. In the entangled state (\ref{MCwf})
which is created by the interference at the beam splitter the
coefficients $c_{n_{01}}(t)$ of significant weight are concentrated in a
small interval around $n_{01}=N_L/2$. That means that the
processes (ii) and (iv)  change the wave function
slightly, but without altering its basic character.

Having established that a well-defined phase still exists in the presence
of pump and collisions, we now turn to its dynamical properties, in 
particular its diffusive behavior. To this end we associate an instantaneous
phase with the Monte Carlo wave function $|\Psi(t)\rangle$ via the relation
\begin{equation}\label{phadef}
\cos^2 [\Phi(t)/2]= \frac{\langle \Psi(t) | (a_0^{\dagger}+b_0^{\dagger})
((a_0 + b_0) | \Psi(t)\rangle}{2 \langle \Psi(t) | a_0^{\dagger}a_0+
b_0^{\dagger}b_0 | \Psi(t)\rangle},
\end{equation}
compare with Eq.~(\ref{cosrel}). Note that Eq.~(\ref{phadef}) accounts 
for the fact that the phase is only well-defined for the interval between 
0 and $\pi$. Note also that the use of other equivalent definitions of
the relative phase, e.~g.\ the maximum of the phase distribution
$|c(\Phi)|^2$, is possible. Eq.~(\ref{phadef}) allows one to numerically
evaluate averages of the form
\begin{equation}\label{Adef}
A_{\tilde{\Phi}}(t) = \langle \cos^2 [\Phi(t)/2]\rangle_{\Phi(t=0)=
\tilde{\Phi}}.
\end{equation}
which measures the diffusion of an ensemble of Monte Carlo wave functions
which all have the same initial relative phase $\bar{\Phi}$. One can also
determine the variance
\begin{equation}\label{Vdef}
V_{\tilde{\Phi}}(t) = \langle \cos^4 [\Phi(t)/2]\rangle_{\Phi(t=0)=
\tilde{\Phi}} - A_{\tilde{\Phi}}^2(t),
\end{equation}
where $\cos^4 [\Phi(t)/2]$ is calculated as the square of $\cos^2
[\Phi(t)/2]$ as inferred from Eq.~(\ref{phadef}). A third quantity of
interest is
\begin{equation}\label{Mdef}
M(t) = \langle \langle \{ \cos^2 [\Phi(t)/2] - \cos^2 [\Phi(t=0)/2] \}^2
\rangle \rangle,
\end{equation}
where the inner set of brackets indicates an average over quantum
trajectories with the same initial $\Phi(t=0)$ and the outer one
averaging over the different values of $\Phi(t=0)$.

The quantities introduced in Eqs.~(\ref{Adef}) -- (\ref{Mdef}) allow one
to study the diffusion properties of the relative phase as defined by
Eq.~(\ref{phadef}). If this definition is meaningful then there should
be a connection between this result and the phase dynamics of an individual
atom laser which, as we have seen in Sec.~III, is well described within
the framework of the linearized fluctuation approach. In particular,
the probability distribution for the phase $\phi_0$ of a single laser is
given to a good approximation by
\begin{equation}\label{phadis}
P(\phi_0,t)=\frac 1 {\sqrt{2\pi\sigma_{33}(t)}} \exp\left\{
-(\phi_0-\bar{\phi}_0)^2/ [2\sigma_{33}(t)]\right\}
\end{equation}
where $\bar{\phi}_0$ is the laser phase at $t=0$ and $\sigma_{33}(t)$ is
given by Eq.~(\ref{defsigma}).

Eq.~(\ref{phadis}) predicts that the time dependence of the quantities 
defined in Eqs.~(\ref{Adef}) -- (\ref{Mdef}) are of the form
\begin{eqnarray}
A_{\tilde{\Phi}}(t) & = & \textstyle{\frac 1 2} \left\{ 1+
\exp[-\sigma_{33}(t)] \cos \tilde{\Phi} \right\}, \nonumber \\
V_{\tilde{\Phi}}(t) & = & \textstyle{\frac 1 8} \large\{ 1 -
2 \exp[-2\sigma_{33}(t)] \cos^2 \tilde{\Phi} + \nonumber \\
& & \exp[-4\sigma_{33}(t)]\cos 2\tilde{\Phi} \large\}, \nonumber \\
M(t) & = & \textstyle{\frac 1 4} \left\{ 1-\exp[-\sigma_{33}(t)] \right\}.
\label{preds}
\end{eqnarray}
where the calculation of $M(t)$ assumes that all initial phase
differences have equal probability.

Fig.~\ref{fig42} compares these predictions to a Monte Carlo evaluation of
Eqs.(\ref{Adef}) -- (\ref{Mdef}). Fig.~\ref{fig42}(a) and (b) show $A_0(t)$,
$V_{0}(t)$ and $V_{\pi/2}(t)$ for the collisionless case and the other
parameters as in Fig. 5. Figure \ref{fig42}(c) depicts $M(t)$ for
$V_{0000}=0.25\kappa_0$. In all cases the behavior of the numerical
results is well described by the expressions (\ref{preds}). A similar
degree of agreement was also found in further examples. These results
indicate that it is indeed possible to associate a definite relative phase 
with single simulations of the interference experiment and that this
heuristically introduced phase behaves as predicted by the description
of the atom laser in terms of stochastic processes \cite{Gar91}.

Experimentally, the phase can be observed (as soon as the correlation
time is much larger than the mean time delay between atom emissions) by
monitoring the ratio of the output intensities at the two beam splitter
ports. Thus, the notion of spontaneously broken symmetry appears not be
the most natural concept to explain these experiments, either.

As to the numerical simulations it should be remarked that $\Phi(t)$ as
defined through Eq.~(\ref{phadef}) never assumes the values 0$^{\circ}$ or
180$^{\circ}$. Instead, it always remains in the region between
approximately 10$^{\circ}$ and 170$^{\circ}$, for the collisionless case
(and the specific parameters of our simulations), and a somewhat narrower
interval in the presence of collisions. In fact, the numerical
calculations of $A_0(t)$ and $V_{0}(t)$ were performed using the actual
minimum value of $\Phi$ as initial value.
For this value $\cos^2 (\Phi/2)$ is still close to 1.
At first sight, this observation might invalidate the description of the
phase dynamics in terms of an unbounded stochastic process. However, this
seems rather to be a shortcoming of the simple definition (\ref{phadef}). It
is more appropriate to think of the phase as having at every instant a
well-localized but not infinitely narrow distribution. Equation
(\ref{phadef}) assigns a definite value to this distribution but may
become inaccurate at the extreme values of $\cos^2[\Phi(t)/2]$. In the
calculations, this is compensated for by the fact that the probability
that $\cos^2[\Phi(t)/2]$ assumes its actual minimum or maximum is
somewhat larger than the probability for intermediate values.
Furthermore, it should be noted that in Fig.~\ref{fig42}(c) the numerical
result assumes quite exactly the predicted long-time value of 0.25. This
justifies thinking of the probability distribution of the relative phase
as being approximately constant.

\section{Summary and conclusions}

Using a linearized fluctuation analysis we have obtained analytical 
estimates for the influence of elastic collisions on the linewidth of a
two-mode atom laser system. In case elastic interactions between
laser mode atoms provide the dominant collision mechanism the power
spectrum is Gaussian and the linewidth scales as
$V_{0000}\sqrt{\bar{n}_0}$ for weak enough collisions. It undergoes a
transition to a Lorentzian, with a linewidth proportional to ${\bar n}_0$,
as $V_{0000}$ increases. A Schawlow-Townes-like behavior of the linewidth
is recovered if collisions between pump and laser mode
atoms are dominant. In that case the lineshape is Lorentzian with a linewidth
proportional to $V_{0101}^2 / \bar{n}_0$. The accuracy of the analytical 
predictions was confirmed by Monte Carlo wave function simulations. 
In addition, it was shown in a numerical study
that the recently proposed atom laser scheme of
Refs.~\cite{GuzMooMey96,MooMey971,MooMey972} might be capable of
producing linewidths of the order of or smaller than the natural
linewidth $\kappa_0$ of the laser mode. In the course of the
numerical study the usefulness of the analytical estimates in the
search for suitable operating regimes became apparent.

In a second aspect of the phase dynamics, we investigated the relative
phase between two interfering atom lasers. This study extends earlier
work on the relative phase of Bose condensates to the realm of open systems.
The analysis of suitably defined phase distribution functions showed that
a physically meaningful relative phase can be ascribed to single runs
of an interference experiment. A close connection with the
phase dynamics of the single atom laser, as described by a Fokker-Planck
equation, was established by studying the diffusion properties of
this relative phase.

\acknowledgements
We have benefited from numerous discussions with E. V. Goldstein and 
M. G. Moore. This work is supported in part by the U.S. Office of
Naval Research Contract No. 14-91-J1205, by the National Science
Foundation Grant PHY95-07639, by the U.S. Army Research Office and by the
Joint Services Optics Program.

\begin{figure}
\caption{Schematic three-mode atom laser scheme.}
\label{fig21}
\end{figure}

\begin{figure}
\caption{Comparison between numerical calculation of the correlation
function (full curve) and the result of Eq.~(\protect\ref{defc0})
(dashed) for parameter values $\Gamma/\kappa_0=15$,
$\kappa_1/\kappa_0=100$, $N=1.85$, and $V_{0000}/\kappa_0=2.5$, $V_{0101}=0$.}
\label{fig31}
\end{figure}

\begin{figure}
\caption{Inverse half-life time $\tau_{1/2}^{-1}$ of the correlation
function as a function of $V_{0000}$ for parameter values
$\Gamma/\kappa_0=15$, $\kappa_1/\kappa_0=10$, $N=20.3$ and $V_{0101}=0$.
Shown are the results of Eq.~(\protect\ref{defc0}) (full curve),
Eq.~(\protect\ref{g00small}) (dashed), Eq.~(\protect\ref{g00large})
(dotted), and numerical calculations ($\bullet$).}
\label{fig32}
\end{figure}

\begin{figure}
\caption{Inverse half-life time $\tau_{1/2}^{-1}$ of the correlation
function as a function of $V_{0101}$ for parameter values
$\Gamma/\kappa_0=15$, $\kappa_1/\kappa_0=10$, $N=20.3$ and $V_{0000}=0$.
Shown are the results of Eq.~(\protect\ref{defc0}) (full curve),
Eq.~(\protect\ref{g01app}) (dashed), and numerical calculations ($\bullet$).}
\label{fig33}
\end{figure}

\begin{figure}
\caption{Examples of phase distributions $|c(\Phi)|^2$: (a) typical atom
laser state for the collisionless case (other parameters as given in the
text); (b) phase state as defined by Eq.~(\protect\ref{phasestate})
with $\Phi=45^{\circ}$ and $N_L=120$; (c) typical atom laser state in
the presence of collisions ($V_{0000}=0.25\kappa_0$).}
\label{fig41}
\end{figure}

\begin{figure}
\caption{Comparison between numerical calculation (full curves, averages
over approx.\ 250 simulations each) and
analytical prediction (dashed) according to Eqs.~(\protect\ref{preds})
for: (a) $A_{0}(t)$ with $V_{0000}=0$; (b) $V_{0}(t)$ and $V_{\pi/2}(t)$
with $V_{0000}=0$; (c) $M(t)$ with
$V_{0000}=0.25\kappa_0$. Other parameters as given in the text.}
\label{fig42}
\end{figure}

\begin{references}

\bibitem{AndTowMie97}
M.~R. Andrews, C.~G. Townsend, H.-J. Miesner, D.~S. Durfee, D.~M. Kurn, and
  W.~Ketterle, Science {\bf 275}, 637 (1997).

\bibitem{MewAndKur97}
M.-O. Mewes, M.~R. Andrews, D.~M. Kurn, D.~S. Durfee, C.~G. Townsend, and
  W.~Ketterle, Phys. Rev. Lett. {\bf 78}, 582 (1997).

\bibitem{WisCol95}
H.~M. Wiseman and M.~J. Collett, Phys. Lett. A {\bf 202}, 246 (1995).

\bibitem{SprPfaJan95}
R.~J.~C. Spreeuw, T.~Pfau, U.~Janicke, and M.~Wilkens, Europhys. Lett.
  {\bf 32}, 469 (1995).

\bibitem{OlsCasDal95}
M.~Olshanii, Y.~Castin, and J.~Dalibard, in {\em Proceedings of the 12th
 International Conference on Laser Spectroscopy}, edited by M.~Inguscio,
 M.~Allegrini, and A.~Lasso (World Scientific, Singapore, 1995).

\bibitem{JanWil96}
U.~Janicke and M.~Wilkens, Europhys. Lett. {\bf 35}, 561 (1996).

\bibitem{MoyHopSav97}
G.~M. Moy, J.~J. Hope, and C.~M. Savage, Phys. Rev. A {\bf 55}, 3631
  (1997).

\bibitem{GuzMooMey96}
A.~Guzman, M.~Moore, and P.~Meystre, Phys. Rev. A {\bf 53}, 977 (1996).

\bibitem{WisMarWal96}
H.~Wiseman, A.~Martins, and D.~Walls,  Quantum Semiclass. Opt. {\bf 8},
  737 (1996).

\bibitem{HolBurGarZol96}
M.~Holland, K.~Burnett, C.~Gardiner, J.~I. Cirac, and P.~Zoller, Phys.
  Rev. A {\bf 54}, R1757 (1996).

\bibitem{MooMey971}
M.~G. Moore and P.~Meystre, Phys. Rev. A {\bf 56}, 2989 (1997).

\bibitem{MooMey972}
M.~G. Moore and P.~Meystre, J. Mod. Opt. {\bf 44}, 1815 (1997).

\bibitem{BalBurSco97}
R.~J. Ballagh, K.~Burnett, and T.~F. Scott, Phys. Rev. Lett. {\bf 78},
  1607 (1997).

\bibitem{Hop97}
J.~J. Hope, Phys. Rev. A {\bf 55}, R2531 (1997).

\bibitem{NarSchWal97}
M.~Naraschewski, A.~Schenzle, and H.~Wallis, Phys. Rev. A {\bf 56}, 603
  (1997).

\bibitem{MoySav97}
G.~M. Moy and C.~M. Savage, (preprint).

\bibitem{AndEnsMat95}
M.H. Anderson, J.R. Ensher, M.R. Matthews, C.E. Wieman, and E.A. Cornell,
  Science {\bf 269}, 198 (1995).

\bibitem{DavMewAnd95}
K.~B. Davis, M.-O. Mewes, M.~R. Andrews, N.~J. van Druten, D.~S. Durfee, D.~M.
  Kurn, and W.~Ketterle, Phys. Rev. Lett. {\bf 75}(22), 3969 (1995).

\bibitem{BraSacTolHul95}
C.~C. Bradley, C.~A. Sackett, J.~J. Tollett, and R.~G. Hulet, Phys. Rev.
  Lett. {\bf 75}, 1687 (1995).

\bibitem{GarZol97}
C.~W. Gardiner and P.~Zoller, Phys. Rev. A {\bf 55}, 2902 (1997).

\bibitem{JakGarZol97}
D.~Jaksch, C.~W. Gardiner, and P.~Zoller, Phys. Rev. A {\bf 56}, 575
  (1997).

\bibitem{GarZolBalDav97}
C.~W. Gardiner, P.~Zoller, R.~J. Ballagh, and M.~J. Davis, Phys. Rev.
  Lett. {\bf 79}, 1793 (1997).

\bibitem{Wis97}
H.~M. Wiseman, Phys. Rev. A {\bf 56}, 2068 (1997).

\bibitem{DumParZolGar92}
R.~Dum, A.~S. Parkins, P.~Zoller, and C.~W. Gardiner, Phys. Rev. A {\bf
  46}, 4382 (1992).

\bibitem{MolCasDal93}
K.~M{\o}lmer, Y.~Castin, and J.~Dalibard, J. Opt. Soc. Am. B {\bf 10},
  524 (1993).

\bibitem{JavYoo96}
J.~Javanainen and S.~M. Yoo, Phys. Rev. Lett. {\bf 76}, 161 (1996).

\bibitem{NarWalSch96}
M.~Naraschewski, H.~Wallis, A.~Schenzle, J.~I. Cirac, and P.~Zoller, Phys.
  Rev. A {\bf 54}, 2185 (1996).

\bibitem{CirGarNar96}
J.~I. Cirac, C.~W. Gardiner, M.~Naraschewski, and P.~Zoller, Phys. Rev. A
  {\bf 54}, R3714 (1996).

\bibitem{WonColWal96}
T.~Wong, M.~J. Collet, and D.~F. Walls, Phys. Rev. A {\bf 54}, R3718
  (1996).

\bibitem{CasDal97}
Y.~Castin and J.~Dalibard, Phys. Rev. A {\bf 55}, 4330 (1997).

\bibitem{JavWil97}
J.~Javanainen and M.~Wilkens, Phys. Rev Lett. {\bf 78}, 4675 (1997).

\bibitem{MeySar91}
P.~Meystre and M.~{Sargent} III,
\newblock {\em Elements of Quantum Optics}
\newblock (Springer-Verlag, Berlin, 1991).

\bibitem{WalMil94}
D.~F. Walls and G.~J. Milburn,
\newblock {\em Quantum Optics}
\newblock (Sprin\-ger-Ver\-lag, Berlin, 1994).

\bibitem{Gar83}
C.~W. Gardiner,
\newblock {\em Handbook of Stochastic Methods}
\newblock (Sprin\-ger-Ver\-lag, Berlin, 1983).

\bibitem{Gar91}
C.~W. Gardiner,
\newblock {\em Quantum Noise}
\newblock (Sprin\-ger-Ver\-lag, Ber\-lin, 1991).

\bibitem{Mol97}
K.~M{\o}lmer, Phys. Rev. A {\bf 55}, 3195 (1997).

\end{references}
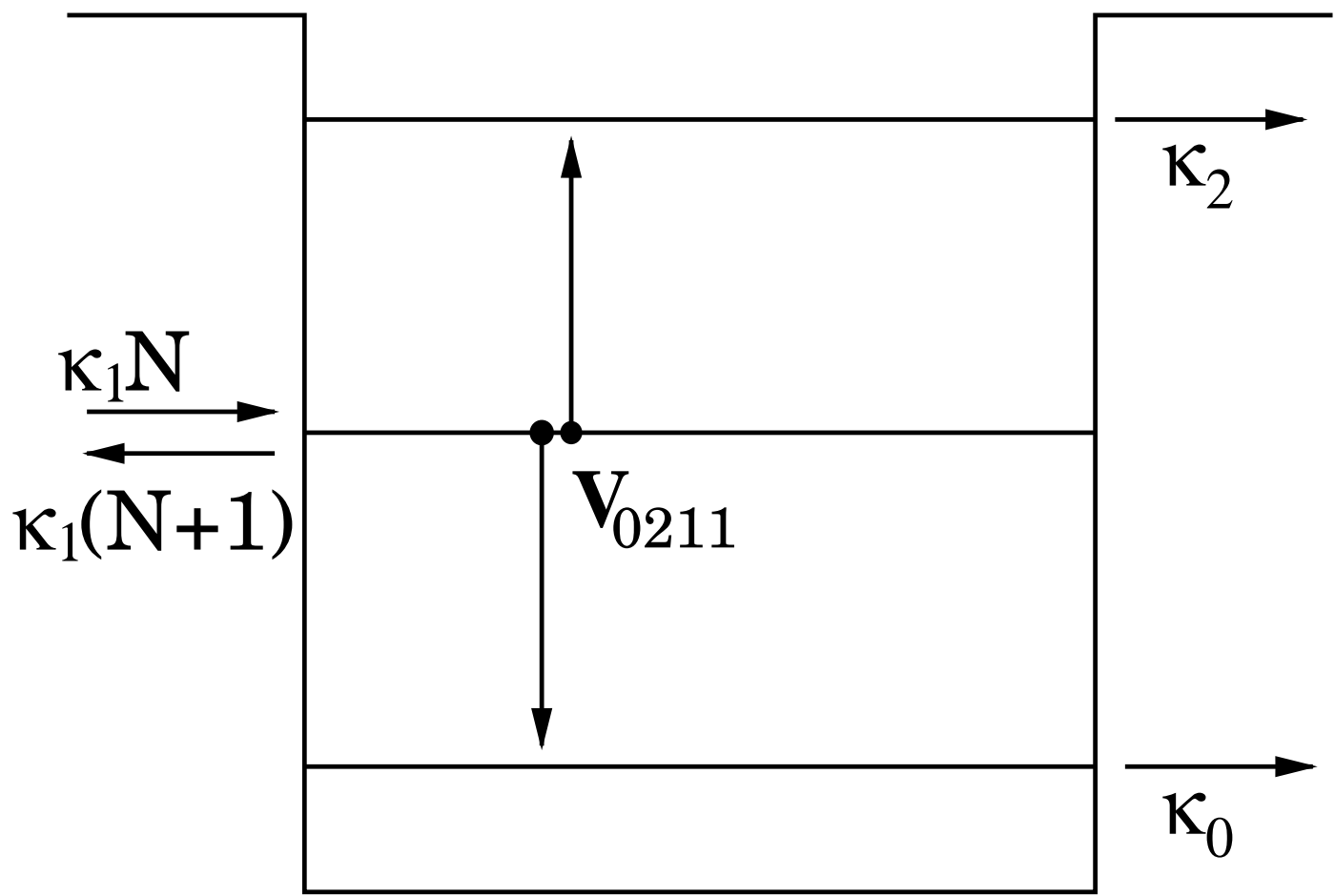
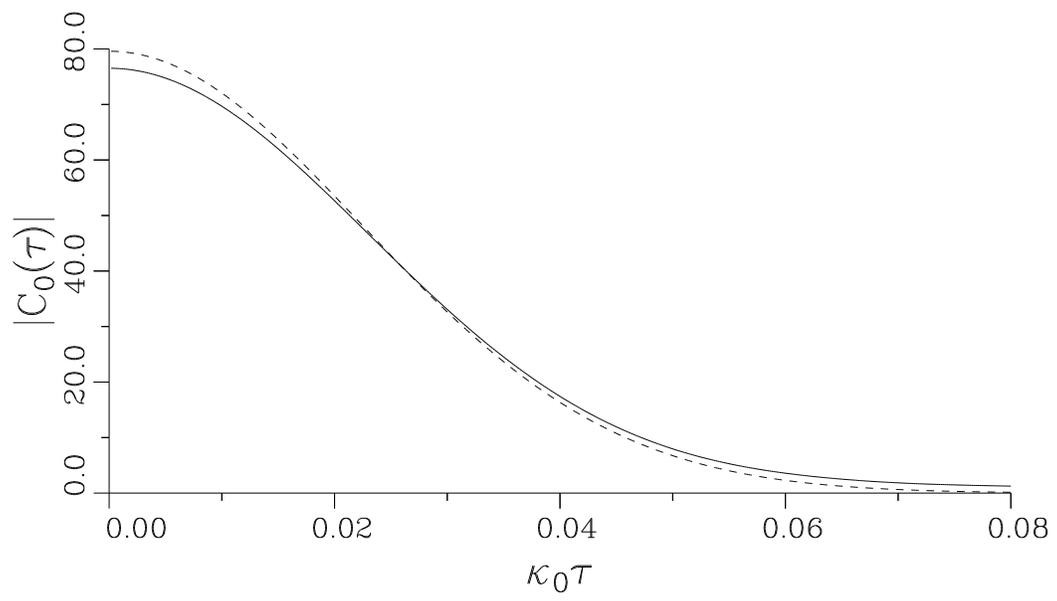
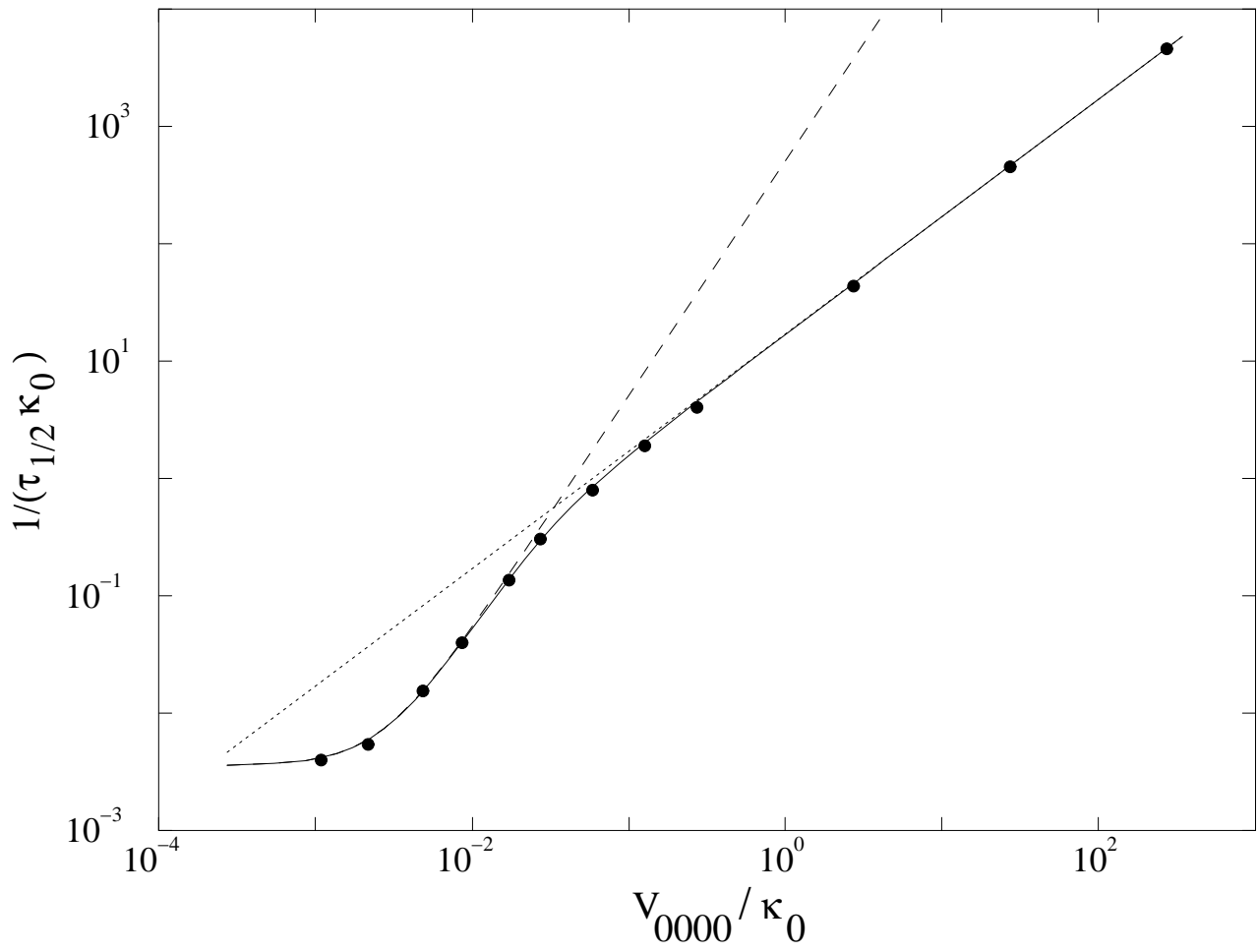
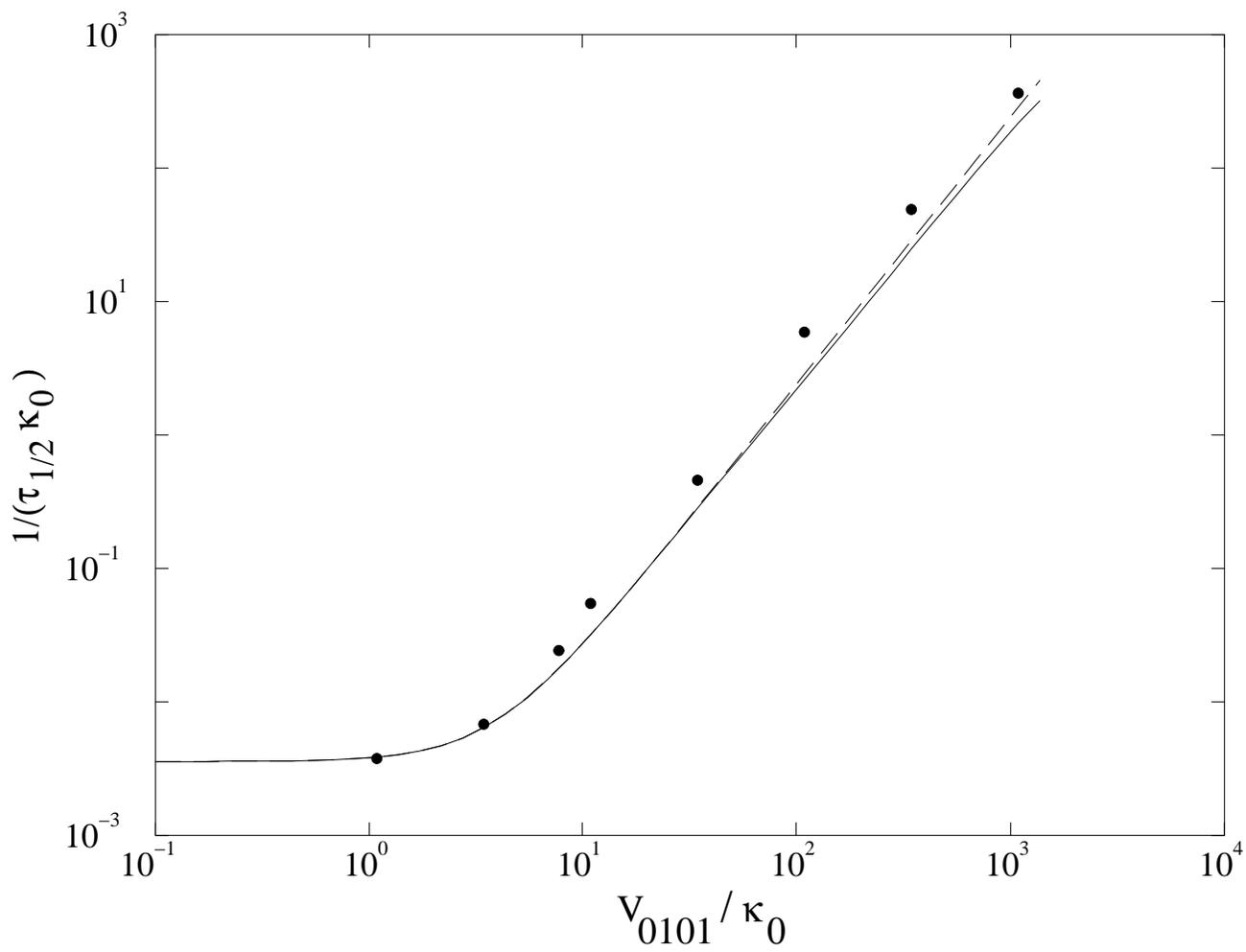
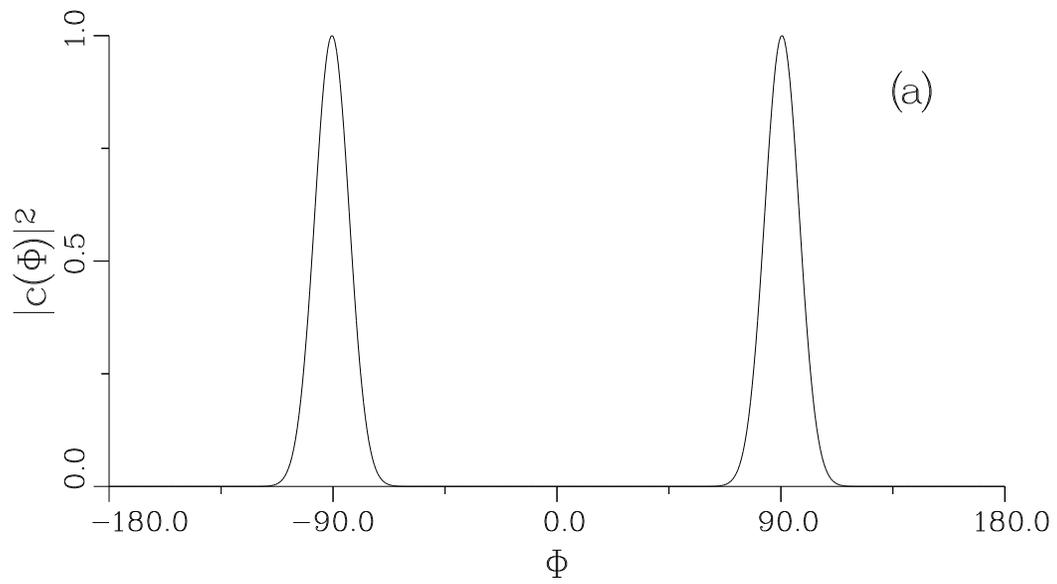
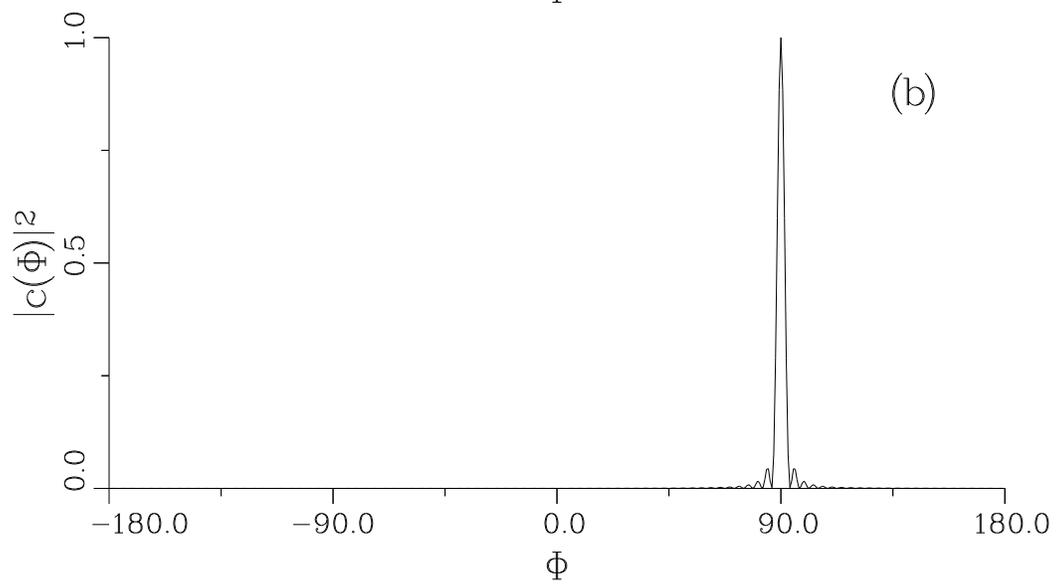
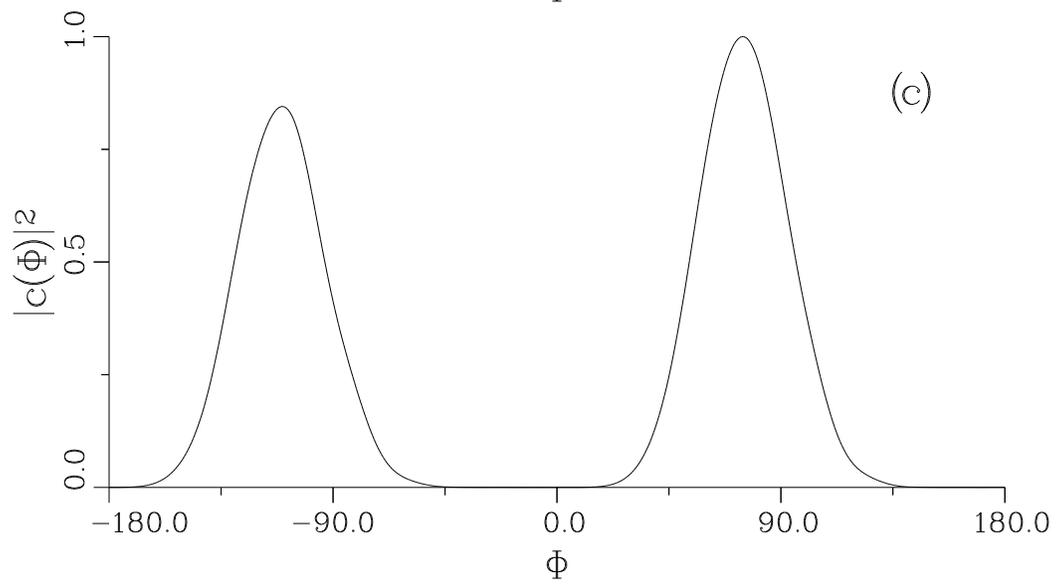
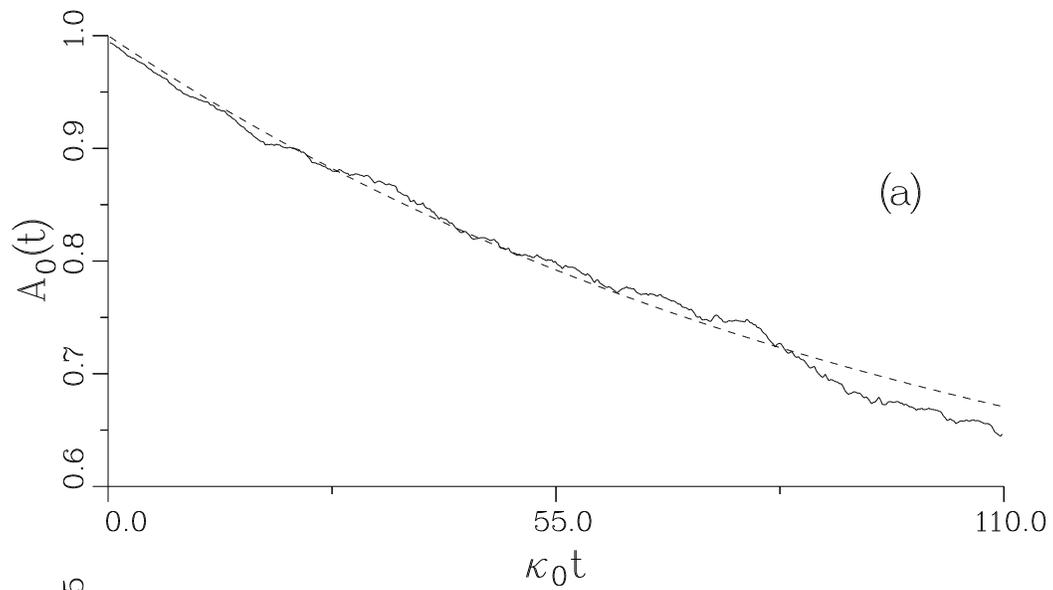
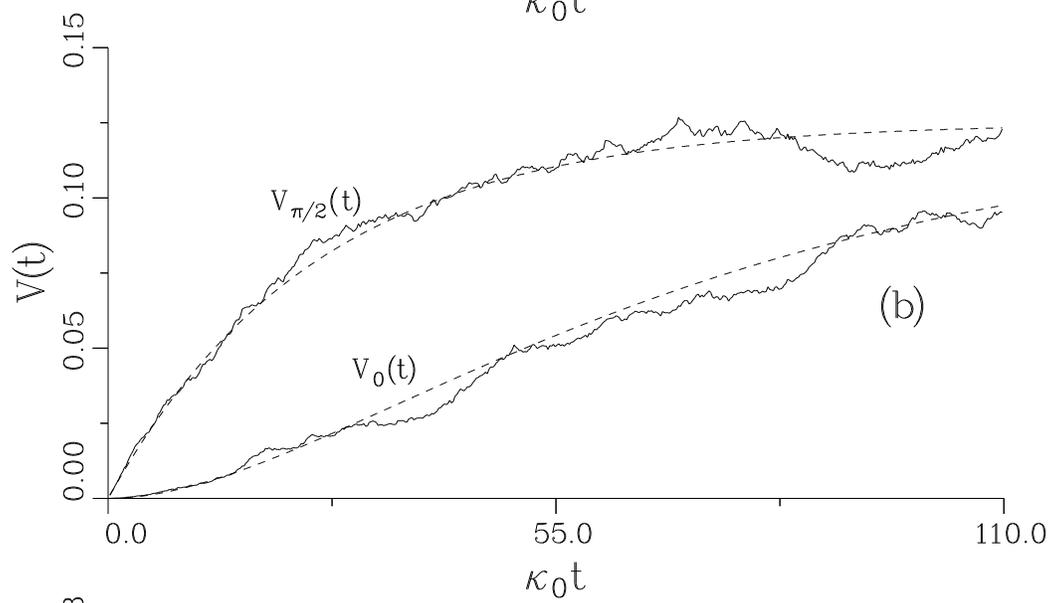
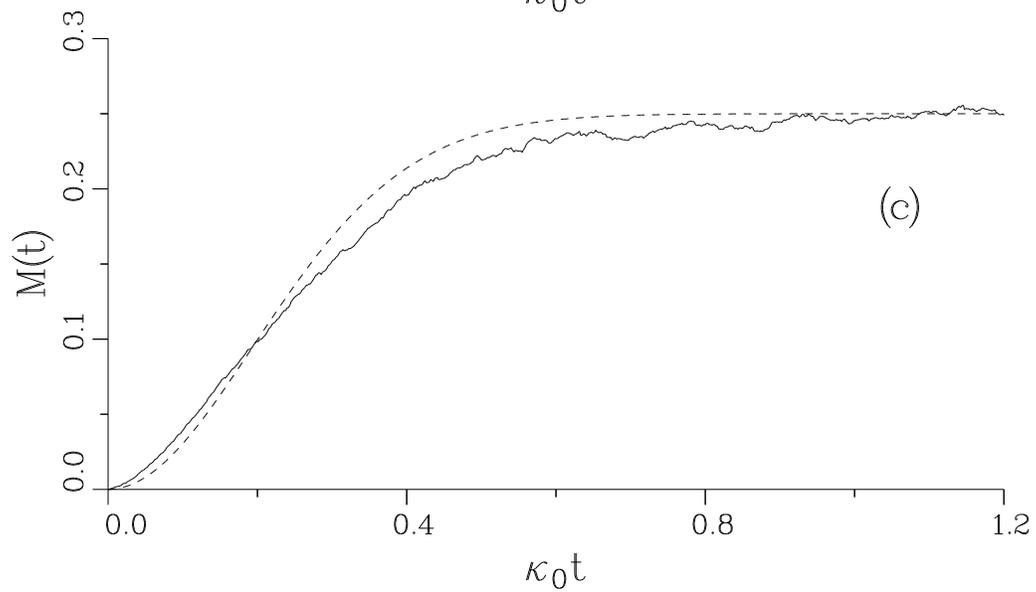
\end{document}